\documentclass[twocolumn,showpacs,showkeys,prl]{revtex4} 

\usepackage{amsmath}

\def\be{\begin{equation}}
\def\ee{\end{equation}}
\def\vec#1{\mathbf{#1}}
\def\del{\partial}
\def\div{\vec{\nabla}\cdot}

\DeclareMathOperator{\e}{e}

\begin{document}

\title{Exciton BCS or BEC state in a semiconductor bilayer system?}
\author{E. H. \surname{Martins Ferreira}}\email{erlon@fisica.ufmg.br}
\author{Tathiana Moreira}
\author{M. C. Nemes}
\affiliation{Departamento de F{\'\i}sica, ICEx, UFMG, \\
P.O.Box 702, 30.161-970, Belo Horizonte, MG, Brazil}

\begin{abstract}
We calculate the off-diagonal long range order (ODLRO) terms of the exciton--exciton correlation function of a semiconductor bilayer system with Coulomb interaction and a transverse magnetic field. We show that the formation of a BEC state is very sensitive to the width of the interaction in momentum space. This dependence is analytically derived and represents the key physical ingredient for the formation (or not) of an exciton condensate state.
\end{abstract}

\maketitle

\section{Introduction}

After the formulation of the pairing model for superconductivity, speculations about the formation of exciton condensates started\cite{keldysh}. Much theoretical and experimental effort was invested in the search of this condensate. However due to the short life-time of optically generated excitons the detection is far from being a simple task. A new approach to this problem  has been then adopted with indirect excitons in semiconductor bilayer systems \cite{eisenstein1,eisenstein2,tutuc} and it has been claimed\cite{eisenstein} that there is a BEC phase in the system. In this last experiment two parallel semiconductor plates with an applied external transverse magnetic field are under investigation. The presence of a magnetic field creates additional energy levels in the system, known as Landau levels, which have a degeneracy proportional to the magnetic flux piercing the plates. Therefore if the magnetic field is strong enough, there will be more available lowest Landau levels (LLL) than the number of electrons, such that the unoccupied levels can be seen as holes in resemblance to the usual holes in semiconductor valence bands. 

Many theoretical papers have dealt with this kind of system as in Refs.~\cite{zhu,doretto,shibata} to cite a few. Nevertheless, a theoretical approach to such system is still a great challenge, as it is for any many--body problem. It is too hard, if not impossible, to determine the ground state of such systems and therefore either approximations or numerical calculations are called for. In the case of an exact numerical treatment of the GS wave function this is restricted to a small number of particles. Here we take a look at the experiment described above and propose simple models for the ground state of that system which allow analytical results and point to the key physical ingredient responsible for the BEC formation. We calculate the off-diagonal long range correlations for the excitons since it characterizes the formation of a condensate. 

\section{Landau levels}

Consider an electron moving on the $x$--$y$ plane under a constant magnetic field $B$ in $z$ direction. One possible gauge for the potential vector $\vec A$ is 
\be \vec A = -By\,{\hat\imath}\ . \ee

Here we find that $\div\vec A=0$, which implies that $\vec p$ and $\vec A$ commute. Writing down the Schr\"odinger equation, we find
\be\begin{split}
-\frac{\hbar^2}{2m}\biggl[\nabla^2 -\frac {ieB}{\hbar c} y \del_x + \frac{e^2B^2}{\hbar^2 c^2}y^2 \biggr]\psi(x,y) & = E\psi(x,y)
\end{split}\ .\ee

We try a solution of the type
\be \psi_{n,k}(x,y) = \e^{ikx}\chi_n(y)\ , \ee
which yields
\be
\biggl[-\frac{\hbar^2}{2m}\frac{\del^2}{\del y^2} + \frac{m\omega_c^2}{2}
(y+ l^2k)^2 \biggr]\chi_n(y)  = E_n\chi_n(y)
\ee
where $\omega_c = eB/(mc)$ is the cyclotron frequency and $l=\sqrt{\hbar c/(eB)}$ is the magnetic length. The solutions are the same of a harmonic oscillator centered at $y_0 = -l^2k$ and are given by
\begin{align} 
\chi_n(y) & = \e^{-(y+l^2k)^2/(2l^2)}H_n\bigl[(y+l^2k)/l\bigr]\ ,\\
 E_n &= \Bigl(n+\frac 12\Bigr)\hbar\omega_c\ . 
\end{align}

Considering only the LLL, i.e., $n=0$, the normalized wave--function is 
\be \psi_k(x,y)=\frac 1{(lL_x\sqrt{\pi})^{1/2}}\e^{ikx} \e^{-(y+l^2k)^2/(2l^2)}\ .\label{eq:ll} \ee

These wave-functions are orthogonal and obey
\begin{subequations}
\begin{align}
\int d^2r\ \psi^*_k(r)\psi_{k'}(r) & = \delta_{k,k'}\ , \\
\sum_k \psi^*_k(r)\psi_{k}(r') & = \delta(r-r')\ .
\end{align}
\end{subequations}
In a finite system, $k$ runs from 1 to $\Omega$, which is the degeneracy of the system. In the present situation of a plate of area $A= L_x\,L_y$, we find $\Omega = B A/\Phi_0$, where $\Phi_0 = h c/e$ is the elementary magnetic flux.

\subsection{Second quantization formalism}

Using second quantization, we define the field operators for the electrons as
\be 
\mathbf{\Psi(r)} = \sum_k \psi_k(r)\mathbf{a_k}\ ,\quad \mathbf{\Psi^\dag(r)} = \sum_k \psi^*_k(r)\mathbf{a_k^\dag}\ ,
\ee 
where $\mathbf{a_k}\ (\mathbf{a^\dag_k})$ are the annihilation (creation) operators in the Fock space. The action of these operators in a Fock state is
\begin{subequations}
\begin{align}
\mathbf{a_k} |n_1 \cdots n_k \cdots n_\Omega\rangle & = n_k |n_1 \cdots 0 \cdots n_\Omega\rangle\ , \\
\mathbf{a^\dag_k} |n_1 \cdots n_k \cdots n_\Omega\rangle & = (1-n_k) |n_1 \cdots 1 \cdots n_\Omega\rangle\ ,
\end{align}
\end{subequations}
and they obey the canonical anticommutation relations
\be \bigl\{\mathbf{a_k, a^\dag_{k'}}\bigr\} = \delta_{kk'}\, ,\  
\bigl\{\mathbf{a_k, a_{k'}}\bigr\} = 0\, ,\   
\bigl\{\mathbf{a_k^\dag, a^\dag_{k'}}\bigr\} = 0\ .
\ee
The total number of particles in such a state is obviously $N = \sum_k n_k$ and this can be written in an operator fashion as
\be \mathbf{N} = \sum_k\mathbf{a^\dag_k a_k} = \int d^2r \mathbf{\Psi^\dag(r)\Psi(r)}\ . \ee  

Let us compute the Fourier transform of the density operator $\mathbf{\rho(r) = \Psi^\dag(r)\Psi(r)}$ for electrons given by
\be \begin{split}
\rho(\mathbf{q}) & = \int d^2 r\e^{-i\mathbf{q\cdot r}}\mathbf{\Psi^\dag(r)\Psi(r)} \\
& = \frac{1}{\pi^{1/2} L_x l}\sum_{k,k'}\int d^2 r\e^{-i(q_xx+q_yy)} \e^{-i(k-k')x} \\
&\quad \times  \exp\biggl\{-\frac{(y+l^2k)^2+(y+l^2k')^2}{2l^2}\biggr\} \ \mathbf{{a^\dag}_k a_{k'}}\\
& = \sum_{k,k'} \frac{\e^{-l^2\Delta k^2/4}}{\pi^{1/2} L_x l}\int d^2 r\e^{-i(q_xx+q_yy)} \e^{i\Delta k x} \\
&\quad \times  \exp\biggl\{-\frac{(y+l^2\bar k)^2}{l^2}\biggr\} \mathbf{{a^\dag}_k a_{k'}}\ .
\end{split} \ee
In the last passage we have defined $\Delta k = k' - k$ and $\bar k = (k+k')/2$. Integrating we get:
\be \begin{split}
\rho(\mathbf{q}) & = \sum_{k,k'} \delta_{q_x,\Delta k}\,\e^{-l^2\Delta k^2/4}  \exp\biggl\{-\frac{l^2q_y^2}4 +iq_yl^2\bar{k}\biggr\} \mathbf{{a^\dag}_k a_{k'}}\\
& = \e^{-l^2q^2/4}\e^{il^2\,q_yq_x/2}\sum_{k} \e^{il^2\,kq_y} \mathbf{{a^\dag}_k a_{k+q_x}}
\end{split} \ee

\subsection{Particle--Hole transformation}

We are initially dealing with electrons in two parallel sheets. However if the number of electrons is smaller than $\Omega$, we can treat the empty levels in one of the sheets (the lower one for instance) as holes. We then define the hole creation operator $\mathbf{\bar \Psi^\dag(r)}$ as the electron destruction one. Thus
\be \mathbf{\bar \Psi^\dag(r)} = \mathbf{\Psi (r)}\ \text{and}\ \mathbf{{b^\dag}_k} = \mathbf{a_k} \ee
Then the Fourier transform of the density operator for holes is similarly
\be \begin{split}
\bar\rho(\mathbf{q}) & =  \int d^2 r\e^{-i\mathbf{q\cdot r}}\mathbf{\bar\Psi^\dag(r)\bar\Psi(r)} \\
& = \e^{-l^2q^2/4}\e^{-il^2\,q_yq_x/2}\sum_{k} \e^{il^2\,kq_y} \mathbf{{b^\dag}_k b_{k-q_x}}
\end{split} \ee

\section{Coulomb interaction}

We are mainly interested in the Coulomb interaction between the particles in different sheets. 
The Fourier transform of the Coulomb potential in 2-D and a separation in $z$ equal $d$ reads
\be V(\mathbf{q}) = \frac{4\pi e^2}{\epsilon_0}\frac{\e^{-qd}}{q}\ ,\ q\neq 0\ . \ee

The coulombic repulsion between electrons in the different sheets can be more conveniently rewritten as a coulombic attraction between electrons and holes by a particle-hole transformation in one of the sheets. Then, this interaction term is given by
\begin{multline}
-\int d^2 r_1 d^2 r_2 \mathbf{\Psi^\dag(r_1)\bar \Psi^\dag(r_2)} V(r_2-r_1)\mathbf{\bar \Psi(r_2)\Psi(r_1)} \\ 
\begin{aligned}
& = -(2\pi)^{-2}\sum_q \rho(\mathbf{q})V(\mathbf{q})\bar \rho(\mathbf{-q}) \\
& =  \sum_{k,k',p} F_{p}\; \mathbf{{a^\dag}_{k} {b^\dag}_{k'} b_{k'+p}a_{k+p}}
\end{aligned}
\end{multline}
where 
\be 
F_p = -\frac{2e}{L_x\epsilon_0}\int dq_y\, \cos(l^2\,q_yq_x) \e^{-l^2q^2/2}  \frac{\e^{-qd}}{q}\biggr\rvert_{q_x=p}
\label{eq:fp}
\ee

Motivated by the model proposed in Ref.~\cite{eisenstein}, we first make  a drastic approach to the interaction potential to be a delta function. From that, we find that $F_p \propto \delta_{p0}$, and we have something similar to a pairing model. The ground state of this model is easy to find and we will use it to calculate the exciton-exciton correlation function. Later, we come back to discuss the real nature of the interaction potential and its implications.

\section{Ground state and correlation function}

For the case that $F_p \propto \delta_{p0}$ (pairing model) we can use the following expression to describe the ground state formed by $n$ excitons:
\be 
| n \rangle = \frac{1}{\sqrt{\mathcal{N}}}\Bigl(\sum_{k} \mathbf{a}^\dag_{k}\mathbf{b}^\dag_{k}\Bigr)^n|0\rangle,
\label{eq:n}\ee 
where $\mathbf{a}^\dag_{k}\ (\mathbf{b}^\dag_{k})$ creates a electron (hole) in the upper (lower) sheet and the normalization being $\mathcal{N}=\Omega!\, n!/(\Omega-n)!$. 

We now look for the general 2-body correlation function as given by 
\begin{multline}
\langle n| \Psi^\dag(r_1)\bar\Psi^\dag(r_2)\bar\Psi(r'_2)\Psi(r'_1)|n\rangle = \\
\bigl \langle n\bigl | \sum_{\substack{k_1,k'_1 \\ k_2, k'_2}} \psi^*_{k_1}(r_1)\mathbf{a}^\dag_{k_1}\, \bar\psi^*_{k_2}(r_2)\mathbf{b}^\dag_{k_2}\,\bar\psi_{k'_2}(r'_2)  
 \mathbf{b}_{k'_2}\,\psi_{k'_1}(r'_1)\mathbf{a}_{k'_1}\bigr|n\bigr \rangle
\end{multline}

With use of Eq.~\eqref{eq:n} and the anticommutation relations of the operators, we can calculate the expected value
\be
\langle n|  \mathbf{a}^\dag_{k_1}\, \mathbf{b}^\dag_{k_2}\, \mathbf{b}_{k'_2}\, \mathbf{a}_{k'_1}|n\rangle = 
C_1 \delta_{k_1 k_2}\delta_{k'_1 k'_2}  + C_1 \delta_{k_1 k'_1}\delta_{k_2 k'_2} 
\label{eq:ev}
\ee
where
\be C_1 = \frac{n(\Omega-n)}{\Omega(\Omega-1)}\quad \text{and} \quad
C_2 = \frac{n(n-1)}{\Omega(\Omega-1)}\ .
\ee

To see if this state is a BEC state, we calculate the exciton--exciton correlation function. This function is given by the creation of a pair electron-hole at one site and its following annihilation at another site, that is
\begin{multline}
\langle n| \Psi^\dag(r)\bar\Psi^\dag(r)\bar\Psi(r')\Psi(r')|n\rangle = \\
\begin{aligned}
& C_1\sum_{k,k'}\psi^*_{k}(r) \bar\psi^*_{k}(r) \psi_{k'}(r') \bar\psi_{k'}(r') \\ 
+\ & C_2\sum_{k,k'} \psi^*_{k}(r) \bar\psi^*_{k'}(r) \psi_{k'}(r') \bar\psi_{k}(r')\ .
\end{aligned}
\end{multline}

First we note that $\bar\psi_k(r) = \psi_k^*(r)$. Then there are only two products to calculate, namely, $\lvert\psi_k(r)\rvert^2$ and $\psi^*_k(r)\psi^*_k(r')$. We begin with the first one. Recalling Eq.~\eqref{eq:ll}, we find 
\begin{equation}
\sum_{k} \bigl\lvert \psi_{k}(r)\bigr\rvert^2 =  \sum_{k} \frac 1{L_x\sqrt{\pi}} \exp\Bigl[-l^2(k+y/l^2)^2\Bigr] 
\end{equation}
We now transform this summation into an integral, with the measure $dk = 2\pi/L_x$. We find then
\be 
\sum_{k} \bigl\lvert \psi_{k}(r)\bigr\rvert^2 \rightarrow \frac 1{2\pi^{3/2} l}\int dk \exp\Bigl[-l^2(k+y/l^2)^2\Bigr] = \frac{1}{2\pi l^2}\ .
\ee

For the second term we find 
\begin{multline}
\sum_{k} \psi^*_{k}(r) \psi^*_{k}(r') = 
\frac {1}{L_x l\sqrt{\pi}}\exp\Bigl[-\frac {(y-y')^2}{4l^2}\Bigr]\\
\times  \sum_{k} \e^{-ik(x-x')}\exp\Bigl[-l^2 \bigl(k+\frac{y+y'}{2\lambda}\bigr)^2\Bigr]\\
\rightarrow  \frac {1}{2\pi l^2}\exp\Bigl(-\frac {|r-r'|^2}{4l^2}\Bigr) \exp\Bigl(-\frac{i}{2\lambda}(y+y')(x-x')\Bigr) \ .
\end{multline}

Thus we get
\begin{multline}
\langle n| \Psi^\dag(r)\bar\Psi^\dag(r)\bar\Psi(r')\Psi(r')|n\rangle = 
\frac {1}{(2\pi l^2)^2}\frac{n}{\Omega(\Omega-1)}\\
\times \biggl\{ (\Omega-n) + (n-1) \exp\Bigl(-\frac {|r-r'|^2}{2l^2}\Bigr) \biggr\} 
\label{eq14}
\end{multline}

Taking the thermodynamic limit of the previous result, with $|r-r'|\rightarrow\infty$ and $n,\Omega\rightarrow\infty$, such that the filling factor $\nu = n/\Omega$ remains finite, we get
\begin{equation}
\langle n| \Psi^\dag(r')\bar\Psi^\dag(r')\bar\Psi(r)\Psi(r)|n\rangle \rightarrow
\frac{\nu(1-\nu)}{(2\pi l^2)^2}
\label{eq29}
\end{equation}

We then conclude that the ODLRO correlation terms remain finite in the thermodynamic limit, which characterizes a BEC state. That means, we would find a BEC if the interaction were sharp enough so that we could approach the potential to a delta function. However, we know that the Coulomb potential is long range, and that changes the scenario completely.

\section{A more realistic approach}

Looking back to Eq.~\eqref{eq:fp}, by plotting $F_p$ for different values of $d/l$, we see that it has a typical width in $p$ of the order of $\Delta p \sim 1/(100 l)$. This means that in this interval we find approximately $\sqrt{\Omega}/250$  values of $p$ which contribute in the summation. In other words, this means that we will find excitons formed by an electron and a hole with different momenta, or simply that the exciton has a spatial `width'. This is also ratified by a numerical solution for a system with a few states.

With this in mind, we must generalize the state for the system in order to take this `exciton width' into account. We then rewrite the state of $n$ excitons with a separation in momentum $p$ as
\be | n, p \rangle = \frac{1}{\sqrt{\mathcal{N}}}\Bigl(\sum_{k} \mathbf{a}^\dag_{k}\mathbf{b}^\dag_{k+p}\Bigr)^n|0\rangle.
\label{eq:np}\ee 

These states are still orthogonal, since
\be \langle m,q| n, p \rangle = \delta_{nm}\delta_{pq}\ . \ee

We can construct now a more general Ansatz for the state of the system as
\be  |\Psi_0\rangle = \sum_{p=0}^{\Omega-1} f_p  |p,n\rangle, \ee
where $\sum_p |f_p|^2 = 1$. We will assume for sake of simplicity that $f_p = \e^{-p^2/2\alpha^2}/\sqrt{(\alpha\sqrt{\pi}+1)/2}$.

To compute the exciton--exciton correlation function in this case, we first calculate the terms
\begin{multline}
\langle p, n| \Psi^\dag(r')\bar\Psi^\dag(r')\bar\Psi(r)\Psi(r)|q, n\rangle = \\
\frac {1}{(2\pi l^2)^2}\frac{n}{\Omega(\Omega-1)}\biggl\{ (n-1) \exp\Bigl[-\frac {(y-y')^2}{2l^2}-\frac{(x-x')^2}{2l^2}\Bigr] \\
+ (\Omega-n)\e^{ip(x'-x)}\e^{-l^2p^2/2} \biggr\} \delta_{pq} \ .
\label{eq:pn}
\end{multline}
Terms not diagonal in $p,q$ yield a contribution of the order of $1/\Omega$, which is negligible in the thermodynamic limit. The same happens to the first term of the above equation. We then have only to consider the following expression
\begin{multline}
\langle \Psi_0| \Psi^\dag(r')\bar\Psi^\dag(r')\bar\Psi(r)\Psi(r)|\Psi_0\rangle \\
\begin{aligned}
& = \frac {1}{(2\pi l^2)^2}\frac{n(\Omega-n)}{\Omega(\Omega-1)} \sum_p \e^{ip(x'-x)} \frac{2\e^{-p^2/\alpha^2}}{\alpha\sqrt{\pi}+1} \e^{-l^2p^2/2} \\
& = \frac {1}{(2\pi l^2)^2}\frac{n(\Omega-n)}{\Omega(\Omega-1)} \sqrt{\frac{2}{\alpha^2l^2+2}}\exp\Bigl[-\frac{\alpha^2(x'-x)^2}{2(\alpha^2l^2+2)}\Bigr]\ .
\end{aligned}
\end{multline} 
	
If $\alpha=0$, we reproduce the previous result of Eq.~\ref{eq29}, as we should expect, then $\alpha=0$ is equivalent to make $f_p=\delta_{p0}$. However, if $\alpha \neq 0$, this term would vanish in the thermodynamic limit.

From the previous result we must conclude that the condensate would be formed for the $|n,p=0\rangle$ state. However, the solely Coulomb interaction is not sharp enough in momentum space to allow for a condensate formation. Based on the experiment related in Ref.~\cite{eisenstein}, we are led to conclude that either there is another interaction between the particles of the different sheets or the state they have found is not a BEC state, but just a collective BCS-like one.

\begin{acknowledgments}
We thank Prof. A.F.R. de Toledo Piza for the helpful discussions. 
This work is supported by Brazilian Financial agencies FAPEMIG, CNPq and FAPESP. 
\end{acknowledgments}

\end{document}